\def\prg#1{\paragraph{#1}}           
            \def\Ra{\Rightarrow}
\def\ni{\noindent}                   \def\pd{\partial}
\def\mb#1{\hbox{\boldmath$#1$}}      \def\grl{{GR$_\L$}}
\def\Leff{\hbox{$\L_{\rm eff}$}}     \def\dis{\displaystyle}
\def\cs{{\scriptstyle\rm CS}}
\def\bull{\raise.25ex\hbox{\vrule height.8ex width.8ex}}
\def\a{\alpha}          \def\b{\beta}            \def\g{\gamma}
\def\d{\delta}               \def\ve{\varepsilon}
\def\th{\theta}         \def\k{\kappa}           \def\l{\lambda}
\def\m{\mu}             \def\n{\nu}              \def\p{\pi}
\def\r{\rho}            \def\s{\sigma}           \def\t{\tau}
\def\vphi{\varphi}      \def\om{\omega}
\def\G{\Gamma}          \def\D{\Delta}           \def\L{\Lambda}
          \def\Om{\Omega}

\def\bA{{\bar A}}       \def\bL{{\bar L}}        \def\bq{{\bar q}}
     \def\tG{{\tilde G}}      \def\tR{{\tilde R}}
\def\cL{{\cal L}}       \def\cH{{\cal H}}        
\def\cM{{\cal M }}      \def\cO{{\cal O}}        \def\cE{{\cal E}}
\def\cK{{\cal K}}
           \def\tH{\tilde H}
\def\bcH{{\bar\cH}}     \def\bcK{{\bar\cK}}      
\def\nn{\nonumber}
\def\be{\begin{equation}}            \def\ee{\end{equation}}
\def\ba#1{\begin{array}{#1}}         \def\ea{\end{array}}
\def\bea{\begin{eqnarray} }          \def\eea{\end{eqnarray} }
\def\lab#1{\label{eq:#1}}            \def\eq#1{(\ref{eq:#1})}
\def\bsubeq{\begin{subequations}} \def\esubeq{\end{subequations}\par\ni}
\def\bitem{\begin{itemize}}          \def\eitem{\end{itemize}}

\documentclass[11pt,twoside]{article}
\catcode`\@=11
\newtoks\@stequation
\def\subequations{\refstepcounter{equation}%
  \edef\@savedequation{\the\c@equation}%
  \@stequation=\expandafter{\theequation}%
  \edef\@savedtheequation{\the\@stequation}%
  \edef\oldtheequation{\theequation}%
  \setcounter{equation}{0}%
  \def\theequation{\oldtheequation\alph{equation}}}
\def\endsubequations{\setcounter{equation}{\@savedequation}%
  \@stequation=\expandafter{\@savedtheequation}%
  \edef\theequation{\the\@stequation}\global\@ignoretrue}
\catcode`\@=12
\def\vsm{\vspace{-8pt}}
\renewcommand{\theequation}{\thesection.\arabic{equation}}
\date{}

\begin{document}

\title{{\bf Anti-de Sitter 3-dimensional gravity with torsion}\footnote{
Invited talk given at {\it Workshop on Quantum Gravity and Noncommutative
Geometry\/}, Universidade Lus\'ofona, Lisbon, 20-23 July 2004,
and at {\it III Summer School in Modern Mathematical Physics\/},
Zlatibor, Serbia, 20-31 August 2004.}}

\author{M. Blagojevi\'c and M. Vasili\'c \vspace{2pt}  \\
{\it Institute of Physics P.O.Box 57, 11001 Belgrade, Serbia}}

\maketitle

\begin{abstract}
Using the canonical formalism, we study the asymptotic symmetries of
the topological 3-dimensional gravity with torsion. In the anti-de
Sitter sector, the symmetries are realized by two independent Virasoro
algebras with classical central charges. In the simple case of the
teleparallel vacuum geometry, the central charges are equal to each
other and have the same value as in general relativity, while in the
general Riemann-Cartan geometry, they become different.

\medskip
\ni{\it keywords:\/} gravity; torsion; asymptotic symmetry.
\end{abstract}

\section{Introduction}

Three-dimensional (3d) gravity has been used as a theoretical
laboratory to test some of the conceptual problems of both classical
and quantum gravity \cite{1,2}. One can identify several particularly
important achievements in the development of these ideas. (a) Brown and
Henneaux demonstrated that, under suitable asymptotic conditions, the
asymptotic symmetry of 3d gravity has an extremely rich structure,
described by two independent canonical Virasoro algebras with classical
central charges \cite{3}. (b) Soon after that, Witten found that general
relativity with a cosmological constant (\grl) can be formulated as a
Chern-Simons gauge theory \cite{4}. The equivalence between gravity and
an ordinary gauge theory was crucial for a deeper understanding of
quantum gravity. (c) Next, the discovery of the black hole solution by
Ba\'nados, Teitelboim and Zanelli had a powerful impact on 3d
gravity \cite{5}. It turned out that the Virasoro algebra of the
asymptotic symmetry plays a central role in our understanding of the
quantum nature of black hole [6-11].

Following a widely spread belief that the dynamics of gravity is to be
described by general relativity, investigations of 3d gravity have been
carried out mostly in the realm of {\it Riemannian\/} geometry.
However, since the early 1990s, the possibility of {\it
Riemann-Cartan\/} geometry has been also explored [12-18]; it is a
geometry in which both the {\it curvature\/} and the {\it torsion\/}
are present as independent geometric characteristics of
spacetime \cite{19,20}. In this way, one expects to clarify the
influence of geometry on the dynamical content of spacetime.

Dynamics of a theory is determined not only by its action, but also by
the asymptotic conditions. The dynamical content of asymptotic
conditions is best seen in topological theories, where the non-trivial
dynamics is bound to exist only at the boundary. General action for
topological 3d gravity in Riemann--Cartan spacetime has been proposed
by Mielke and Baekler \cite{12,13}. This model is our starting point for
exploring the structure of 3d gravity with torsion. In particular,
we shall investigate

\newpage
\bitem
\item[\bull] the existence of the black hole with torsion, and \vsm
\item[\bull] the asymptotic structure of 3d gravity with torsion.
\eitem
We restrict ourselves to the anti-de Sitter (AdS) sector of the theory,
with negative effective cosmological constant. For a particular choice
of parameters, the Mielke-Baekler action leads to the {\it
teleparallel\/} (Weizenb\"ock) geometry in vacuum \cite{21,22,20},
defined by the requirement of vanishing curvature, which we choose as
the simplest framework for studying the influence of torsion on the
spacetime dynamics.

The paper is  organized as follows. In Sect. 2 we introduce
Riemann--Cartan spacetime as a general geometric arena for 3d gravity
with torsion, and discuss the teleparallel description of gravity in
vacuum. In Sect. 3 we construct the teleparallel black hole solution.
Then, in Sect. 4, we introduce the concept of asymptotically AdS
configuration, and show that the related asymptotic symmetry is the
same as in general relativity---the conformal symmetry. In the next
section, the gauge structure of the theory is incorporated into the
canonical formalism by investigating the Poisson bracket (PB) algebra
of the asymptotic generators. The asymptotic symmetry is characterized
by two independent canonical Virasoro algebras with classical central
charges, the values of which are the same as in Riemannian spacetime of
general relativity. In Sect. 6 we discuss the general case of
Riemann-Cartan geometry, and show that the related classical central
charges are different. Finally, Sect. 7 is devoted to concluding
remarks.

Our conventions are given by the following rules: the Latin
indices refer to the local Lorentz frame, the Greek indices refer
to the coordinate frame; the first letters of both alphabets
$(a,b,c,...;$ $\a,\b,\g,...)$ run over 1,2, the middle alphabet
letters $(i,j,k,...;\m,\n,\l,...)$ run over 0,1,2; the signature
of spacetime is $\eta=(+,-,-)$; totally antisymmetric tensor
$\ve^{ijk}$ and the related tensor density $\ve^{\m\n\r}$ are both
normalized so that $\ve^{012}=1$.

\section{Riemann-Cartan gravity in 3d}

Theory of gravity with torsion can be formulated as Poincar\'e
gauge theory (PGT), with an underlying spacetime structure described by
Riemann-Cartan geometry \cite{19,20}.

\prg{PGT in brief.} The basic gravitational variables in PGT are the
triad $b^i=b^i{_\m}dx^\m$ and the Lorentz connection
$A^{ij}=A^{ij}{}_\m dx^\m$ (1-forms). Metric tensor $g$ is not an
independent variable, it is defined in terms of the triad field:
$g=b^i\otimes b^j\,\eta_{ij}$, where $\eta_{ij}=(+,-,-)$. The field
strengths corresponding to the gauge potentials $b^i$ and $A^{ij}$
are the torsion $T^i$ and the curvature $R^{ij}$ (2-forms). In 3d, we
can simplify the notation by introducing the duals of $A^{ij}$ and
$R^{ij}$: $\om_i=-\frac{1}{2}\,\ve_{ijk}A^{jk},
~R_i=-\frac{1}{2}\,\ve_{ijk}R^{jk}$. Gauge symmetries of the theory
are local translations and local Lorentz rotations, parametrized by
$\xi^\m$ and $\th^i$, respectively:
\bea
&&\d_0 b^i{_\m} = -\ve^i{}_{jk}b^j{}_\m\th^k
      -(\pd_\m\xi^\r)b^i{_\r}-\xi^\r\pd_\r b^i{}_\m        \nn\\
&&\d_0\om^i{_\m} = -\nabla_\m\th^i
     -(\pd_\m\xi^\r)\om^i{}_\r-\xi^\r\pd_\r\om^i{}_\m\, ,  \lab{2.1}
\eea
where $\nabla_\m\th^i=\pd_\m\th^i+\ve^i{}_{jk}\om^j{_\m}\th^k$ is the
covariant derivative of $\th^i$. The related field strengths, the
torsion and the curvature, are given by the expressions
\bea
&&R^i=d\om^i + \frac{1}{2}\,\ve^i{}_{jk}\om^j\om^k
     \equiv\frac{1}{2}R^i{}_{\m\n}dx^\m dx^\n\, ,          \nn\\
&&T^i=db^i + \ve^i{}_{jk}\om^j b^k
     \equiv\frac{1}{2}\,T^i{}_{\m\n}dx^\m dx^\n\, ,        \lab{2.2}
\eea
where wedge product signs are omitted for simplicity.

In PGT, the triad and the connection are related to each other by the
{\it me\-tri\-ci\-ty condition\/}: $\nabla\mb{g}=0$. The geometric
interpretation of the connection implies a useful identity, which
relates Lorentz connection $A$ and (Riemannian) Levi-Civita
connection $\D$:
\be
A_{ijk}=\D_{ijk}+K_{ijk}\, ,                               \lab{2.3}
\ee
where $K_{ijk}=-(T_{ijk}+T_{kij}-T_{jki})/2$ is the contortion.

\prg{Topological action.} In genaral, gravitational dynamics is
defined by Lagrangians which are at most quadratic in field
strengths. Omitting the quadratic terms, Mielke and Baekler
proposed a {\it topological\/} model for 3d gravity \cite{12,13},
with the action
\bsubeq\lab{2.4}
\be
I=aI_1+\L I_2+\a_3I_3+\a_4I_4+I_M\, ,                      \lab{2.4a}
\ee
where
\bea
&&I_1=2\int b^iR_i\, ,                                     \nn\\
&&I_2=-\frac{1}{3}\,\int\ve_{ijk}b^ib^jb^k\, ,             \nn\\
&&I_3=\int\left(\om^id\om_i
          +\frac{1}{3}\ve_{ijk}\om^i\om^j\om^k \right)\, , \nn\\
&&I_4=\int b^iT_i\, ,                                      \lab{2.4b}
\eea
\esubeq
and $I_M$ is a matter action. The first term with $a=1/16\pi G$ is
the usual Einstein-Cartan action, the second term is a cosmological
term, $I_3$ is a Chern-Simons action for the connection, and $I_4$ is
an action of the translational Chern-Simons type. The Mielke-Baekler
model can be thought of as a generalization of \grl\ ($\a_3=\a_4=0$)
to a topological gravity theory in Riemann-Cartan spacetime.

\prg{Field equations.} Variation of the action with respect to
triad and connection yields the field equations:
\bea
&&\ve^{\m\n\r}\left[aR_{i\n\r}+\a_4T_{i\n\r}
  -\L\ve_{ijk}b^j{_\n}b^k{_\r}\right]=\t^\m{_i}\, ,        \nn\\
&&\ve^{\m\n\r}\left[\a_3R_{i\n\r}+ aT_{i\n\r}
  +\a_4\ve_{ijk}b^j{_\n}b^k{_\r}\right]=\s^\m{_i}\, ,      \nn
\eea
where $\t$ and $\s$ are the matter energy-momentum and spin tensors,
respectively. For our purposes---to explore exact vacuum solutions
and the asymptotic structure---it is sufficient to consider the field
equations in vacuum, where $\t=\s=0$.  For $\a_3\a_4-a^2\ne 0$, these
equations take the simple form
\bsubeq\lab{2.5}
\bea
&& T_{ijk}=A\ve_{ijk}\, ,                                  \lab{2.5a}\\
&& R_{ijk}=B\ve_{ijk}\, ,                                  \lab{2.5b}
\eea
\esubeq
where
$$
A=\frac{\a_3\L+\a_4 a}{\a_3\a_4-a^2}\, ,\qquad
B=-\frac{(\a_4)^2+a\L}{\a_3\a_4-a^2}\, .
$$
Thus, the vacuum configuration is characterized by constant torsion and
constant curvature.

Using the PGT identity \eq{2.3}, one can express the Riemann-Cartan
curvature $R^{ij}{}_{\m\n}(A)$ in terms of its Riemannian piece
$\tR^{ij}{}_{\m\n}\equiv\tR^{ij}{}_{\m\n}(\D)$ and the contortion.
This geometric identity, combined with the field equations \eq{2.5},
leads to
\be
\tR^{ij}{} _{\m\n}= -\Leff(b^i{_\m} b^j{_\n}-b^i{_\n} b^j{_\m})\, ,
\qquad \Leff=B-\frac 14 A^2\, ,                            \lab{2.6}
\ee
where $\Leff$ is the effective cosmological constant. Equation
\eq{2.6} can be considered as an equivalent of the second field
equation \eq{2.5b}. Looking at \eq{2.6} as an equation for the
metric, one concludes that our spacetime is maximally
symmetric \cite{23}:
\bitem
\item[] $\Leff<0\quad\Ra\quad$ anti-de Sitter sector, \vsm
\item[] $\Leff>0\quad\Ra\quad$ de Sitter sector.
\eitem

\prg{Teleparallelism in vacuum.} There are two interesting special
cases of the general Mielke-Baekler model.
\bitem
\item[$-$] For $\a_3=\a_4=0$, the torsion vanishes and the vacuum
geometry becomes {\it Riemannian\/}, with the field equations
$T_{ijk}=0$, $R_{ijk}=B\ve_{ijk}$. This case defines \grl, and
corresponds to Witten's choice \cite{4}.  \vspace{-7pt}
\item[$-$] for $(\a_4)^2+a\L=0$, the curvature vanishes, and the
vacuum geometry is {\it teleparallel\/}: $T_{ijk}=A\ve_{ijk}$,
$R_{ijk}=0$ \cite{16}. The vacuum field equations are ``geometrically
dual" to those of \grl.
\eitem
Having in mind our intention to study the influence of torsion on the
gravitational dynamics, we restrict our attention to the teleparallel
geometry as the {\it simplest\/} framework involving torsion. In this
case, the field equations take the form
\bsubeq\lab{2.7}
\bea
&& T_{ijk}=A\ve_{ijk}\, ,                                  \lab{2.7a}\\
&& \tR^{ij}{} _{\m\n}=
   -\Leff(b^i{_\m} b^j{_\n}-b^i{_\n} b^j{_\m})\, ,         \lab{2.7b}
\eea
\esubeq
where the effective cosmological constant becomes negative:
$$
\Leff=-\frac{1}{4}A^2\equiv -\frac{1}{\ell^2}<0\, .
$$
Although the general structure of spacetime, in the presence of
matter, is described by Riemann-Cartan geometry, the field equations
\eq{2.7} imply that the effective dynamics in vacuum is
tele\-pa\-rallel.

Since the field equations do not involve $\a_3$ ($A=-\a_4/a$), we can
take, without loss of generality, that $\a_3=0$. In summary, we adopt
the following restrictions on parameters: $\a_3=0$, $\a_4=-2a/\ell$,
$\L=-4a/\ell^2$ (and $a=1/16\pi G$), whereupon the action \eq{2.4}
reduces to
\be
I= a\int d^3x\ve^{\r\m\n}\left[  b^i{_\r}\left(R_{i\m\n}
   -\frac{1}{\ell}T_{i\m\n}\right) +\frac{4}{3\ell^2}\,
    \ve_{ijk}b^i{_\r}b^j{_\m}b^k{_\n}\right]\, .           \lab{2.8}
\ee

\section{Exact vacuum solutions}
\setcounter{equation}{0}

We start our study of 3d gravity with torsion by an analysis of the
exact classical solutions in vacuum. Our search for the exact solutions
is based on the following procedure:
\bitem
\item[$-$] For a given $\Leff$, find a solution of Eq. \eq{2.7b} for the
metric. This step is very simple, since the form of the metric for
maximally symmetric 3d spaces is well known \cite{23}.  \vsm
\item[$-$] Given the metric, find a solution for the triad, such that
$g=b^i\otimes b^j\,\eta_{ij}$. \vsm
\item[$-$] Finally, use Eq. (2.7a) to determine the connection $\om^i$.
\eitem

\prg{Teleparallel black hole.} Equation \eq{2.7b} represents the
Riemannian condition for maximal symmetry, and it has a well known
solution for the metric---the BTZ black hole \cite{5}. In the static
coordinates $x^\m=(t,r,\vphi)$ ($0\le\vphi<2\pi$), the BTZ metric has
the form (in units $4G=1$):
\bea
&&ds^2=N^2dt^2-N^{-2}dr^2-r^2(d\vphi+N_\vphi dt)^2\, ,     \nn\\
&&N^2=\left(-2m+\frac{r^2}{\ell^2}+\frac{J^2}{r^2}\right)\,,
  \qquad N_\vphi=\frac{J}{r^2}\, .                         \lab{3.1}
\eea
As we shall see later, the parameters $m$ and $J$ define the
conserved charges---energy and angular momentum. Given the metric
\eq{3.1}, the triad field can be chosen in the simple, ``diagonal"
form,
\bsubeq\lab{3.2}
\be
b^0=Ndt\, ,\qquad b^1=N^{-1}dr\, ,\qquad
b^2=r\left(d\vphi+N_\vphi dt\right)\, ,                    \lab{3.2a}
\ee
and the related connection is obtained by solving \eq{2.7a}:
\be
\om^0=Ndx^-\, ,\qquad
  \om^1=N^{-1}\left(\frac{1}{\ell}+\frac{J}{r^2}\right)dr\, ,\qquad
\om^2= -\left(\frac{r}{\ell} -\frac{J}{r}\right)dx^-\, ,   \lab{3.2b}
\ee
\esubeq
where $x^\pm=t/\ell\pm\vphi$. Equations \eq{3.2} define the {\it
teleparallel\/} black hole \cite{16}. For Riemann-Cartan black hole,
see Refs. 15 and 18.

\prg{Teleparallel AdS solution.} In Riemannian geometry with negative
$\L$, there is a general solution with maximal number of Killing
vectors (the solutions of the Killing equation $\d_0 g_{\m\n}=0$),
which is called the anti-de Sitter solution, AdS$_3$. Locally, it can
be obtained from \eq{3.1} by the replacement $J=0$, $2m=-1$. Although
AdS$_3$ and the black hole are locally isometric, they are globally
distinct \cite{5,23}.

Similarly, in the teleparallel geometry, the general, maximally
symmetric solution defines the {\it teleparallel\/} AdS$_3$. It can be
obtained from the black hole \eq{3.2} by the same replacement ($J=0$,
$2m=-1$):
\bea
&&b^0=fdt\, ,\hspace{35pt} b^1=f^{-1}dr\, ,
             \qquad  b^2=rd\vphi\, ,                       \nn\\
&&\om^0=fdx^-\, ,\qquad  \om^1=\frac{1}{\ell f}dr\, ,
             \qquad\, \om^2= -\frac{r}{\ell}dx^-\, ,       \lab{3.3}
\eea
with $f^2=1+r^2/\ell^2$.

One should stress that in Riemann-Cartan geometry, one can define the
gene\-ra\-lized isometries by the requirements $\d_0 b^i{_\m}=0$,
$\d_0\om^i{_\m}=0$, which differ from the Killing equation in
Riemannian geometry. When applied to the teleparallel AdS solution
\eq{3.3}, these requirements define six independent solutions
$\xi_{(k)}$ and $\th_{(k)}$ ($k=1,\dots,6$) for the allowed $\xi$ and
$\th$, displayed in Appendix A of Ref. 16. The related symmetry group
is the six-parameter AdS group $SO(2,2)$.

\section{Asymptotic conditions}
\setcounter{equation}{0}

Spacetime outside localized matter sources is described by the vacuum
solutions of the field equations. Thus, matter has no influence on
the local properties of spacetime in the source-free regions, but it
can change its global properties. The global properties can be
expressed geometrically by the symmetry properties of the
{\it asymptotic configurations\/}, which are, on the other hand,
closely related to the gravitational conservation laws.

Returning to 3d gravity with $\Leff<0$, let us note that maximally
symmetric AdS solution \eq{3.3} has the role analogous to the role of
Minkowski space in the $\Leff=0$ case. Following this analogy, we could
choose that all dynamical variables approach to the configuration
\eq{3.3} in such a way, that the resulting asymptotic symmetry is
$SO(2,2)$, the maximal symmetry of \eq{3.3}. However, the important
black hole geometries would be thereby excluded, as they are not
$SO(2,2)$ invariant. Such a situation motivates us to introduce the
concept of the {\it AdS asymptotic behaviour\/}, defined by the
following requirements \cite{3,24}:
\bitem
\item[(a)] the asymptotic conditions should include the black hole
           geometries;   \vsm
\item[(b)] they should be invariant under the action of the AdS group
           $SO(2,2)$;    \vsm
\item[(c)] the asymptotic symmetries should have well defined canonical
           generators.
\eitem

\prg{AdS asymptotics.} According to (a), the asymptotic form of the
black hole con\-fi\-gu\-ration, defined by Eqs. \eq{3.2a} and
\eq{3.2b}, should be included in the set of asymptotic states we are
searching for. The second requirement (b) is realized by acting on
the black hole solution with all possible $SO(2,2)$ transformations,
and demanding the resulting configurations to belong to our set of
asymptotic states.

The procedure just described leads to the following asymptotic form
of the triad field:
\bsubeq\lab{4.1}
\be
b^i{_\m}=\left( \ba{ccc}
       \dis\frac{r}{\ell}+\cO_1   & O_4  & O_1   \\
       \cO_2 & \dis\frac{\ell}{r}+\cO_3  & O_2   \\
       \cO_1 & \cO_4                     & r+\cO_1
                \ea
         \right)
\equiv \left( \ba{ccc}
       \dis\frac{r}{\ell} & 0    & 0  \\
       0 & \dis\frac{\ell}{r}    & 0  \\
       0 & 0                     & r
              \ea
       \right)+B^i{_\m}   \, .                             \lab{4.1a}
\ee
The real meaning of the result is clarified by the following remark.
The set of all $SO(2,2)$ transformations is defined by six pairs
$(\xi_{(k)},\th_{(k)})$, hence, strictly speaking, the family of
black hole triads obtained by the action of these transformations is
parametrized by six real parameters, say $\s_i$. The meaning of
\eq{4.1a} is slightly different: any $c/r^n$ term it supposed to be
of the form $c(t,\vphi)/r^n$, i.e. constants $c=c(\s_i)$ of the six
parameter family are promoted to functions $c(t,\vphi)$. This is the
simplest way to characterize the asymptotic behaviour of $b^i{_\m}$.
The triad family \eq{4.1a} generates the Brown-Henneaux asymptotic
form of the metric \cite{3}, but clearly, it is not uniquely
determined by it.

In a similar manner, we can use the torsion equation of motion (2.7a)
to obtain the asymptotic form of the connection:
\be
\om^i{_\m}=\left( \ba{ccc}
    \dis\frac{r}{\ell^2}+\cO_1 & \cO_2 &\dis -\frac{r}{\ell}+\cO_1 \\
    \cO_2 & \dis\frac{1}{r}+\cO_3 & \cO_2                  \\
    \dis-\frac{r}{\ell^2}+\cO_1 & \cO_2 & \dis\frac{r}{\ell}+\cO_1
                  \ea
           \right)
     \equiv\left( \ba{ccc}
    \dis\frac{r}{\ell^2} & 0 &\dis -\frac{r}{\ell}         \\
    0 & \dis\frac{1}{r}  & 0                               \\
    \dis-\frac{r}{\ell^2} & 0 & \dis \frac{r}{\ell}
                  \ea
           \right)+\Om^i{_\m} \, .                         \lab{4.1b}
\ee
\esubeq

All the higher order terms $B^i{_\m}$ and $\Om^i{_\m}$ are considered
to be arbitrary and independent of each other. One can verify that the
asymptotic conditions \eq{4.1} are indeed invariant under the action of
the AdS group $SO(2,2)$.

\prg{Asymptotic symmetries.} We are now going to examine the
symmetries of the above asymptotic configurations, i.e. to find out
the subset of the gauge transformations which leaves the set of
asymptotic states \eq{4.1} invariant. The parameters of these
transformations are determined by the relations
\bsubeq\lab{4.2}
\bea &&\ve^{ijk}\th_j
b_{k\m}-(\pd_\m\xi^{\r})b^i{_\r}
        -\xi^\r\pd_\r b^i{}_\m=\d_0 B^i{_\m}\, ,           \lab{4.2a}\\
&&-\pd_\m\th^i+\ve^{ijk}\th_j\om_{k\m}-(\pd_\m\xi^\r)\om^i{_\r}
        -\xi^{\r}\pd_\r\om^i{}_{\m}=\d_0\Om^i{_\m}\, .     \lab{4.2b}
\eea
\esubeq
Acting on a specific field satisfying \eq{4.1}, these transformations
change the form of the non-leading terms $B^i{_\m}$, $\Om^i{_\m}$. One
should stress that the symmetry transformations defined in this way
differ from the usual symmetries, which act according to the rule
$\d_0 b^i{_\m}=0$, $\d_0\om^i{_\m}=0$.

We find the restricted gauge parameters as follows \cite{16}. The
symmetric part of \eq{4.2a} multiplied by $b_{i\n}$ (six relations)
yields
\bsubeq\lab{4.3}
\bea
&&\xi^0=\ell\left[ T
  +\frac{1}{2}\left(\frac{\pd^2 T}{\pd t^2}\right)
              \frac{\ell^4}{r^2}\right] +\cO_4\, ,         \lab{4.3a}\\
&&\xi^2=S-\frac{1}{2}\left(\frac{\pd^2 S}{\pd\vphi^2}\right)
              \frac{\ell^2}{r^2}+\cO_4\, ,                 \lab{4.3b}\\
&&\xi^1=-\ell\left(\frac{\pd T}{\pd t}\right)r+\cO_1\, ,   \lab{4.3c}
\eea
where the functions $T(t,\vphi)$ and $S(t,\vphi)$ satisfy the
conditions
$$
\frac{\pd T}{\pd\vphi}=\ell\frac{\pd S}{\pd t}\, ,\qquad
\frac{\pd S}{\pd\vphi}=\ell\frac{\pd T}{\pd t}\, .         \eqno(4.4)
$$
These equations define the two-dimensional conformal group at
large distances, in accordance with the Brown-Henneaux result for
\grl\ \cite{3}.

The remaining three components of \eq{4.2a} are used to
determine $\th^i$:
\bea
&&\th^0=-\frac{\ell^2}{r}T_{,02}+\cO_3\, ,                 \nn\\
&&\th^2=\frac{\ell^3}{r}T_{,00}+\cO_3\, ,                  \nn\\
&&\th^1=T_{,2}+\cO_2\, ,                                   \lab{4.3d}
\eea
\esubeq
\setcounter{equation}{4}
while the conditions \eq{4.2b} produce no new limitations on the
parameters.

Rewriting Eqs. (4.4) in the form $\pd_\pm(T\mp S)=0$, with
$2\pd_\pm=\ell\pd_0 \pm\pd_2$, we find that the general solution is
given by
\be
T+S=f(x^+)\, ,\qquad  T-S=g(x^-)\, ,                       \lab{4.5}
\ee
where $f$ and $g$ are two arbitrary, periodic functions.

Parameters $(T,S)$ define the {\it conformal\/} symmetry in the
asymptotic region of our spacetime. In addition to the conformal
transformations, the complete gauge group defined by \eq{4.3} contains
also the {\it residual\/} (or pure) gauge transformations,
characterized by the higher order terms that remain after imposing
$T=S=0$. As we shall see, the residual gauge transformations do not
contribute to the values of the conserved charges (their generators
vanish weakly), and consequently, they can be ignored in our discussion
of the conserved charges. This is effectively done by introducing the
improved definition of the {\it asymptotic symmetry\/}:
\bitem
\item[$-$] the asymptotic symmetry group is defined as the factor
group of the gauge group determined by \eq{4.3}, with respect to the
residual gauge group.
\eitem

In conclusion, the asymptotic behaviour \eq{4.1} defines the most
general configuration space of the theory that respects our
requirements (a) and (b) formulated at the beginning of this section.
In order to verify the status of the last requirement (c), it is
necessary to explore the canonical structure of the theory.

\section{Canonical structure of the asymptotic symmetry}
\setcounter{equation}{0}

After having introduced the notion of the asymptotic symmetry group,
we now continue with the related canonical analysis. We construct the
improved form of the general gauge generator, assuming the asymptotic
conditions \eq{4.1} and \eq{4.3}, and prove the conservation of the
corresponding charges; then, we investigate the canonical algebra of
asymptotic generators \cite{16}.

\prg{Hamiltonian and constraints.}\hspace{-9pt} Starting from the
definition of the canonical momenta $(\pi_i{^\m},\Pi_i{^\m})$,
corresponding to the Lagrangian variables $(b^i{_\m},\om^i{_\m})$, we
use the Dirac procedure for the constrained dynamical systems to
explore the dynamical structure of the theory.

The primary constraints are of the form:
\bea
&&\phi_i{^0}\equiv\pi_i{^0}\approx 0\, , \hspace{85pt}
   \Phi_i{^0}\equiv\Pi_i{^0}\approx 0\, ,                  \nn\\
&&\phi_i{^\a}\equiv\pi_i{^\a}
  +\frac{2a}{\ell}\ve^{0\a\b}b_{i\b}\approx 0\, , \qquad
   \Phi_i{^\a}\equiv\Pi_i{^\a}-2a\ve^{0\a\b}b_{i\b}\approx 0\,.\nn
\eea
Up to an irrelevant divergence, the total Hamiltonian reads
\be
\cH_T=b^i{_0}\cH_i+\om^i{_0}\cK_i
         +u^i{_0}\pi_i{^0}+v^i{_0}\Pi_i{^0} \, ,           \lab{5.1}
\ee
where
\bea
&&\cH_i=-a\ve^{0\a\b}\left(R_{i\a\b}-\frac{2}{\ell}T_{i\a\b}
  +\frac{4}{\ell^2}\,\ve_{ijk}b^j{_\a}b^k{_\b}\right)
  -\nabla_\b\phi_i{^\b}
  +\frac{2}{\ell}\ve_{imn}b^m{_\b}\phi^{n\b}\, ,           \nn\\
&&\cK_i=-a\ve^{0\a\b}\left(T_{i\a\b}
  -\frac{2}{\ell}\ve_{ijk}b^j{_\a}b^k{_\b}\right)
  -\nabla_\b\Phi_i{^\b}-\ve_{imn}b^m{_\b}\phi^{n\b} \, .   \nn
\eea
The constraints are classified as follows: ($\pi_i{^0},
\Pi_i{^0},\cH_i,\cK_i$) are first class, while
($\phi_i{^\a},\Phi_i{^\a}$) are second class.

\prg{Canonical generators.} Gauge symmetries of a dynamical system
are best described by the canonical generator, which is
constructed using the general method of Castellani \cite{25}.
Expressed in terms of the conveniently chosen parameters $\xi^\m$
and $\th^i$, the gauge generator is given by
\bea
&& G=-G_1-G_2\, ,                                          \nn\\
&&G_1\equiv\dot\xi^\r\left(b^i{_\r}\pi_i{^0}+\om^i{_\r}\Pi_i{^0}\right)
  +\xi^\r\left[b^i{_\r}\bcH_i +\om^i{_\r}\bcK_i
  +(\pd_\r b^i{_0})\pi_i{^0}+(\pd_\r\om^i{_0})\Pi_i{^0}\right]\,,\nn\\
&&G_2\equiv\dot\th^i\Pi_i{^0}
  +\th^i\left[\bcK_i-\ve_{ijk}\left( b^j{_0}\pi^{k0}
  +\om^j{_0}\,\Pi^{k0}\right)\right]\, .                   \lab{5.2}
\eea
Here, the time derivatives $\dot b^i{_\m}$ and $\dot\om^i{_\m}$
are shorts for $u^i{_\m}$ and $v^i{_\m}$, respectively, and the
integration symbol $\int d^2x$ is omitted in order to simplify the
notation.

The transformation law of the fields, defined by the Poisson bracket
$\d_0\phi\equiv \{\phi\,,G\}$, is in complete agreement with the
gauge transformations \eq{2.1} {\it on shell\/}.

\prg{Asymptotics of the phase space.} To complete the analysis of the
asymptotic structure of phase space, we need to define the behaviour
of momentum variables at large distances. Our procedure is based on
the following general principle:
\bitem
\item[$-$] the expressions that vanish on-shell should have an
arbitrarily fast asymptotic decrease, as no solution of the field
equations is thereby lost.
\eitem
Applied to the primary constraints of the theory, this principle
gives the asymptotic behaviour of the momenta $\p_i{^\m}$ and
$\Pi_i{^\m}$. The same principle can be also applied to the secondary
constraints and the true equations of motion, producing a refined
form of the original asymptotic conditions.

We are now ready to discuss the impact of the adopted boundary
conditions on the form of the canonical generator.

\prg{The improved generator.} The canonical symmetry generators act
on dynamical variables via the PB operation, which is defined in terms
of functional derivatives. Hence, the phase-space functionals
representing the gauge generators must have {\it well defined
functional derivatives\/}. Our general gauge generator $G$ does not
meet this requirement, but the problem is corrected by adding suitable
boundary terms \cite{26}.

The improved gauge generator $\tG$ is found to have the following
form:
\bsubeq\lab{5.3}
\bea
&&\tG= G +\G \, ,                                          \nn\\
&&\G=-\int_0^{2\p}\!d\vphi\Big( \ell T\cE^1+S\cM^1\Big)\, ,\lab{5.3a}
\eea
where the integration goes over the circle at infinity (the boundary
of the spatial section of spacetime), and
\bea
&&\cE^{\a}=2a\,\ve^{\a\b0}\left(\om^0{_\b}
  +\frac{1}{\ell}\,b^2{_\b}
  -\frac{1}{\ell}\,b^0{_\b}\right)b^0{_0}\, ,              \nn\\
&&\cM^{\a}=-2a\,\ve^{\a\b 0}\left(\om^2{_\b}
  +\frac{1}{\ell}\,b^0{_\b}
  -\frac{1}{\ell}\,b^2{_\b}\right)b^2{_2}\, .              \lab{5.3b}
\eea
\esubeq
The adopted asymptotic behaviour guarantees finiteness of $\G$,
hence, $\tG$ is a well defined generator (finite and differentiable
functional). Thus, the requirement (c), defined at the beginning of
section 4, is automatically satisfied.

As we can see, the surface term $\G$ depends only on the parameters
$(T,S)$, and not on the higher order terms in \eq{4.3}. Thus, it is
only the {\it asymptotic} generators that have non-trivial surface
terms, or charges. On the other hand, the {\it residual\/} gauge
generators are characterized by vanishing $\G$, and can only have
zero charges \cite{3,24}.

Two special cases of the improved generator \eq{5.5} are of particular
importance: the time translation generator $\tG[\xi^0]$, and the
spatial rotation generator $\tG[\xi^2]$. For $\xi^0=1$ and $\xi^2=1$,
the corresponding surface terms have the meaning of energy and angular
momentum, respectively,:
\be
E_0=\int_0^{2\p}\!\cE^1d\vphi\, ,\qquad
M_0=\int_0^{2\pi}\!\cM^1 d\vphi\, .                        \lab{5.4}
\ee

\prg{Canonical algebra.} We now wish to find the PB algebra of the
improved ge\-ne\-ra\-tors, which contains important information
regarding the symmetry structure of the asymptotic dynamics.

Introducing the notation $G'\equiv G[T',S']$, $G''\equiv G[T'',S'']$,
and so on, the PB algebra is found to have the form
\bsubeq\lab{5.5}
\be
\left\{\tG'',\,\tG'\right\} =\tG''' + C''' \,.             \lab{5.5a}
\ee
where the parameters $T'''$, $S'''$  are determined by the relations
\bea
&&T'''=T'S''_{,\,2}-T''S'_{,\,2}
       +S'T''_{,\,2}-S''T'_{,\,2} \, ,                     \nn\\
&&S'''=S'S''_{,\,2}-S''S'_{,\,2}
       +T'T''_{,\,2}-T''T'_{,\,2}\, ,                      \nn
\eea
and $C'''\equiv C[T',S'\,;\,T'',S'']$ is the {\it central term\/} of
the canonical algebra:
\be
C'''=2a\ell\int d\vphi\left(S''_{,\,2}T'_{,\,22}-
    S'_{,\,2}T''_{,\,22}\right)\,.                         \lab{5.5b}
\ee
\esubeq

\prg{Conservation laws.} Direct calculation based on the PB
algebra \eq{5.5} shows that the asymptotic generator $\tG[T,S]$ is
conserved:
\be
\frac{d}{dt}\tG[T,S]=\frac{\pd\tG}{\pd t}
   +\left\{\tG,\,\tH_T\right\}\approx 0\, .                \lab{5.6}
\ee
This also implies the conservation of the boundary term $\G$.

To test the obtained result, we calculate the values of all the
conserved charges for the BTZ black hole solution \eq{3.2}. Recalling
that $a=1/16\p G=1/4\pi$ (in units $4G=1$), we obtain
$$
E_0({\rm black~hole})= m\, ,\qquad M_0({\rm black~hole})= J\, .
$$
Thus, the black hole parameters $m$ and $J$ have the meaning of
energy and angular momentum, respectively. One can also show that
there are no other independent conserved charges \cite{16}.

\prg{Central charge.} The PB algebra \eq{5.7} can be brought to a
more familiar form by using the representation in terms of Fourier
modes. Indeed, after introducing
$$
2L_n=-\tG[T=S=e^{inx^+}]\, ,\qquad 2\bL_n=-\tG[T=-S=e^{inx^-}]\, .
$$
the canonical algebra takes the form of two independent Virasoro
algebras with classical central charges:
\bea
&& \left\{L_n,\,L_m\right\}=
   -i(n-m)L_{n+m}-2\pi i\,a\ell\,n^3\d_{n,-m}\,,           \nn\\
&& \left\{\bL_n,\,\bL_m\right\}=
   -i(n-m)\bL_{n+m}-2\pi i\,a\ell\,n^3\d_{n,-m}\,.         \nn\\
&& \left\{L_n,\,\bL_m\right\}=0 \, .                       \lab{5.7}
\eea
Upon the redefinition of the zero modes, $L_0\to L_0+\pi a\ell$,
$\bL_0\to \bL_0+\pi a\ell$, we obtain the standard form of the
Virasoro algebras. Using the string theory normalization of the
central charge, we have
\be
c_1=c_2=12\cdot 2\pi a\ell=\frac{3\ell}{2G}\, .            \lab{5.8}
\ee
Thus, two central charges in the teleparallel theory coincide with
each other, and with the Brown--Henneaux central charge, defined in
Riemannian \grl.

The form \eq{5.7} of the asymptotic algebra shows that central term
for the AdS subgroup, generated by $(L_{-1},L_0,L_1)$,
$(\bL_{-1},\bL_0,\bL_1)$, vanishes. This is a consequence of the fact
that the AdS subgroup is an exact symmetry of the vacuum
\eq{3.3} \cite{3}.

\section{Central charges in Riemann-Cartan gravity}
\setcounter{equation}{0}

We now return to the general Riemann-Cartan action \eq{2.4} to
discuss the form of the corresponding asymptotic symmetry \cite{18}.

Starting from the Chern-Simons Lagrangian,
$\cL_\cs(A)=A^idA_i+\frac{1}{3}\ve_{ijk}A^iA^jA^k$,
and introducing the new variables $A^i=\om^i+qb^i$ and
$\bA^i=\om^i+\bq b^i$, with $q\ne \bar q$, one can derive the
important identity
\bsubeq
\bea
\k_1\cL_\cs(A)-\k_2\cL_\cs(\bA)&=&
    2a b^iR_i -\frac{1}{3}\L\ve_{ijk}b^ib^jb^k             \nn\\
  &&+\a_3\cL_\cs(\om)+\a_4 b^iT_i + ad(b^i\om_i)\, ,       \lab{6.1a}
\eea
where
\bea
&&a=\k_1q-\k_2\bq\, ,\qquad \L=-(\k_1q^3-\k_2\bq^3)\, ,    \nn\\
&&\a_3=\k_1-\k_2\, ,\qquad \a_4=\k_1q^2-\k_2\bq^2\, .      \lab{6.1b}
\eea
\esubeq
Comparing with \eq{2.4}, we can rewrite this identity in the
form:
\be
\k_1I_\cs[A]-\k_2I_\cs[\bA]=I_G+ a\int d(b^i\om_i) \, ,    \lab{6.2}
\ee
where $I_G$ denotes the gravitational action in \eq{2.4}. The
additional surface integral on the right hand side is just a
correction which improves the differentiability of $I_G$ under the
boundary conditions \eq{4.1}. Thus, Riemann-Cartan gravity in 3d can
be formulated as a Chern-Simons gauge theory. The role of the
coefficients $\k_1$ and $\k_2$ is to define central charges of the
asymptotic symmetry.

Since $q\ne\bar q$, one can easily find the inverse of \eq{6.1b}:
\bea
&&q=-\frac{A}{2}+\frac{1}{\ell}\, ,\qquad
    \bq=-\frac{A}{2}-\frac{1}{\ell}\, ,                    \nn\\
&&\k_1-\k_2=\a_3\, ,\qquad
    \k_1+\k_2=\ell\left(a+\frac{A}{2}\,\a_3\right)\, ,     \lab{6.3}
\eea
where $\ell=\sqrt{-\Leff}\,$ is real. These relations clarify the
role of four parameters appearing in the action \eq{2.4}. In
particular, combining \eq{6.2} and \eq{6.3} one concludes that the
gravitational theory with $\a_3\ne 0$ has conformal symmetry with two
{\it different\/} central charges:
\be
c_{1,2}= 12\cdot 4\pi\k_{1,2}
       = 24\pi\left[a\ell
         +\a_3\left(\frac{A\ell}{2} \pm 1\right)\right]\, .\lab{6.4}
\ee
In the complementary sector with $\a_3=0$ [\grl\ and our teleparallel
theory \eq{2.8}], the central charges $c_1$ and $c_2$ are equal to
each other.

\section{Concluding remarks}

\indent

$\bull$~ 3d gravity with torsion, defined by Eq. \eq{2.8}, possesses
the teleparallel black hole solution, a generalization of the
Riemannian BTZ black hole.

$\bull$~ Assuming the AdS asymptotic conditions, the canonical asymptotic
symmetry is realized by two commuting Virasoro algebras with central
extensions:   \\
$-$ in \grl\ and in the teleparallel theory: $c_1=c_2=3\ell/2G$,  \\
$-$ in Riemann-Cartan theory: $c_1\ne c_2$.

\section*{Acknowledgments}

This work was supported by the Serbian Science foundation, Serbia.

\end{document}